\documentclass[twoside]{dis09}
\usepackage[latin1]{inputenc}
\usepackage[dvips]{graphicx,epsfig,color}
\usepackage{wrapfig,rotating}
\usepackage{amssymb,amsmath,array}

\begin{document}
\title{New Charmonium-like States at B-Factories}

\author{Thomas Kuhr\\ on behalf of the BaBar and Belle Collaborations
%
%
\vspace{.3cm}\\
%
Institut f\"ur Experimentelle Kernphysik - KIT \\
Wolfgang-Gaede-Str.\ 1, 76131 Karlsruhe - Germany
%
}

\maketitle

\begin{abstract}
Since the discovery of the $X(3872)$ in 2003 the B-factory experiments BaBar and Belle 
have been driving forces in the spectroscopy of new charmonium-like states.
In this article the latest results on the decays $X(3872) \rightarrow D^0\bar{D}^{*0}$
and $X(3872) \rightarrow \psi\gamma$, where $\psi$ stands for $J/\psi$ or $\psi(2S)$, 
are presented and analyses of $Z(4430)^+ \rightarrow \psi\pi^+$ and 
$Z(4050/4250)^+ \rightarrow \chi_{c1}\pi^+$ are discussed.
\end{abstract}

\section{Introduction}
Several new states have been discovered in the last few years\ \cite{Godfrey:2008nc},
most of them at B-factories.
The B-factories provide a clean environment to search for and study these new particles either in
$B$ meson decays or in the production in $e^+e^-$ reactions with initial state radiation.
In this article the recent analyses of new states in $B$ decays are presented.
Both experiments, BaBar and Belle, together have now a total sample of 1.3 billion $B\bar{B}$
pairs available.

The oldest member of the family of new charmonium-like states is the $X(3872)$,
discovered by Belle in 2003\ \cite{X3872:Belle}.
Despite several measurements of its properties at B-factories and at the Tevatron
the nature of this particle is still unknown.
It does not fit into the predicted spectrum of conventional chamonium states.
Since its mass is very close to the $D^0\bar{D}^{*0}$ threshold the hypothesis
of a $D^0\bar{D}^{*0}$ molecule is considered\ \cite{Tornqvist:2004qy}.
Another explanation is a bound state of a diquark and an anti-diquark\ \cite{Maiani:2004vq}.

It is also possible, though unlikely, that the $X(3872)$ is a $c\bar{c}$ state 
and our theoretical models fail to describe it.
Therefore the observation of the $Z(4430)^+$ was received with great interest.
As it is charged it can not be a conventional charmonium state.

\section{$\mathbf{X(3872) \rightarrow D^0\bar{D}^{*0}}$}
\begin{figure}
\centering
\begin{minipage}{0.60\textwidth}
\includegraphics[width=\textwidth]{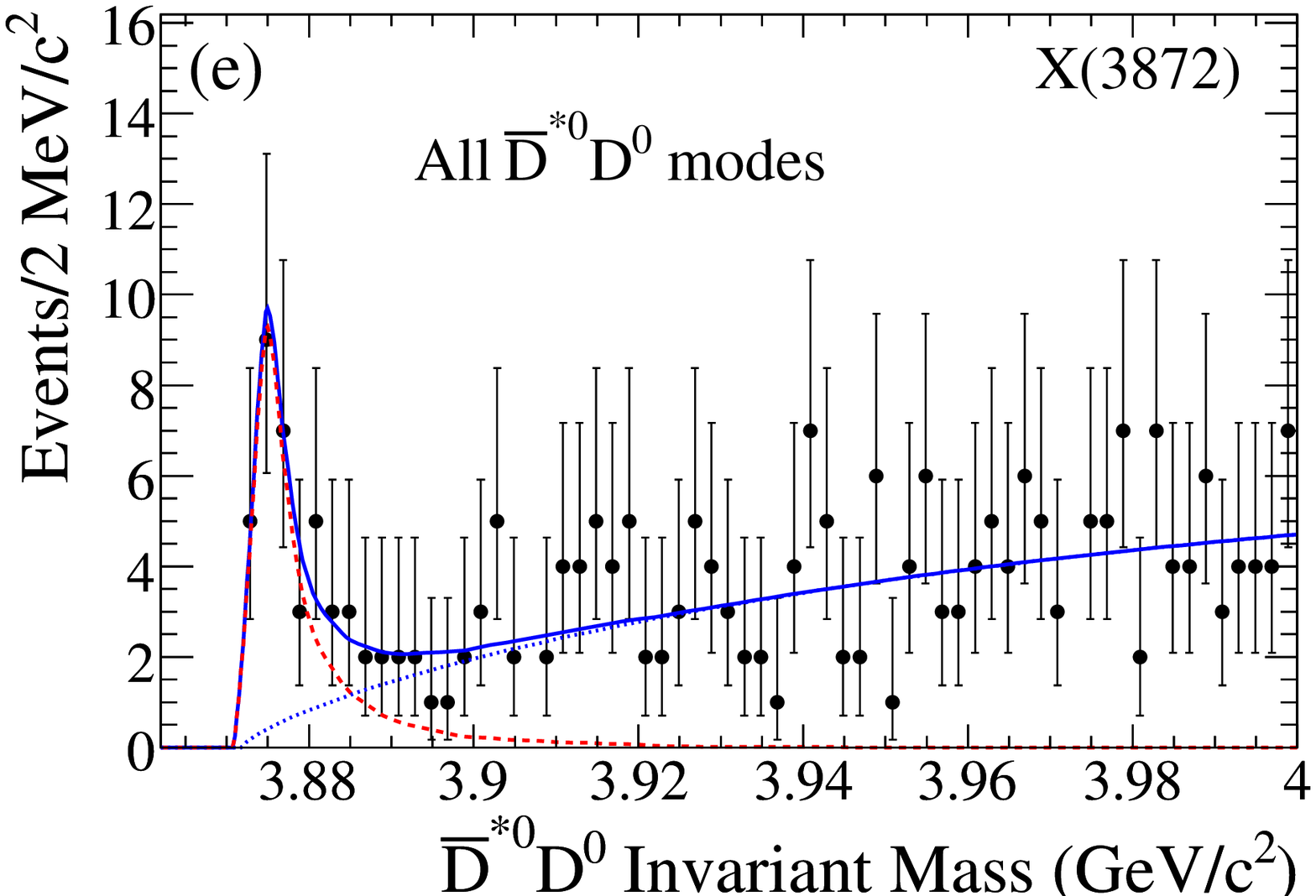}
\end{minipage}
\begin{minipage}{0.38\textwidth}
\includegraphics[clip=true,width=\textwidth]{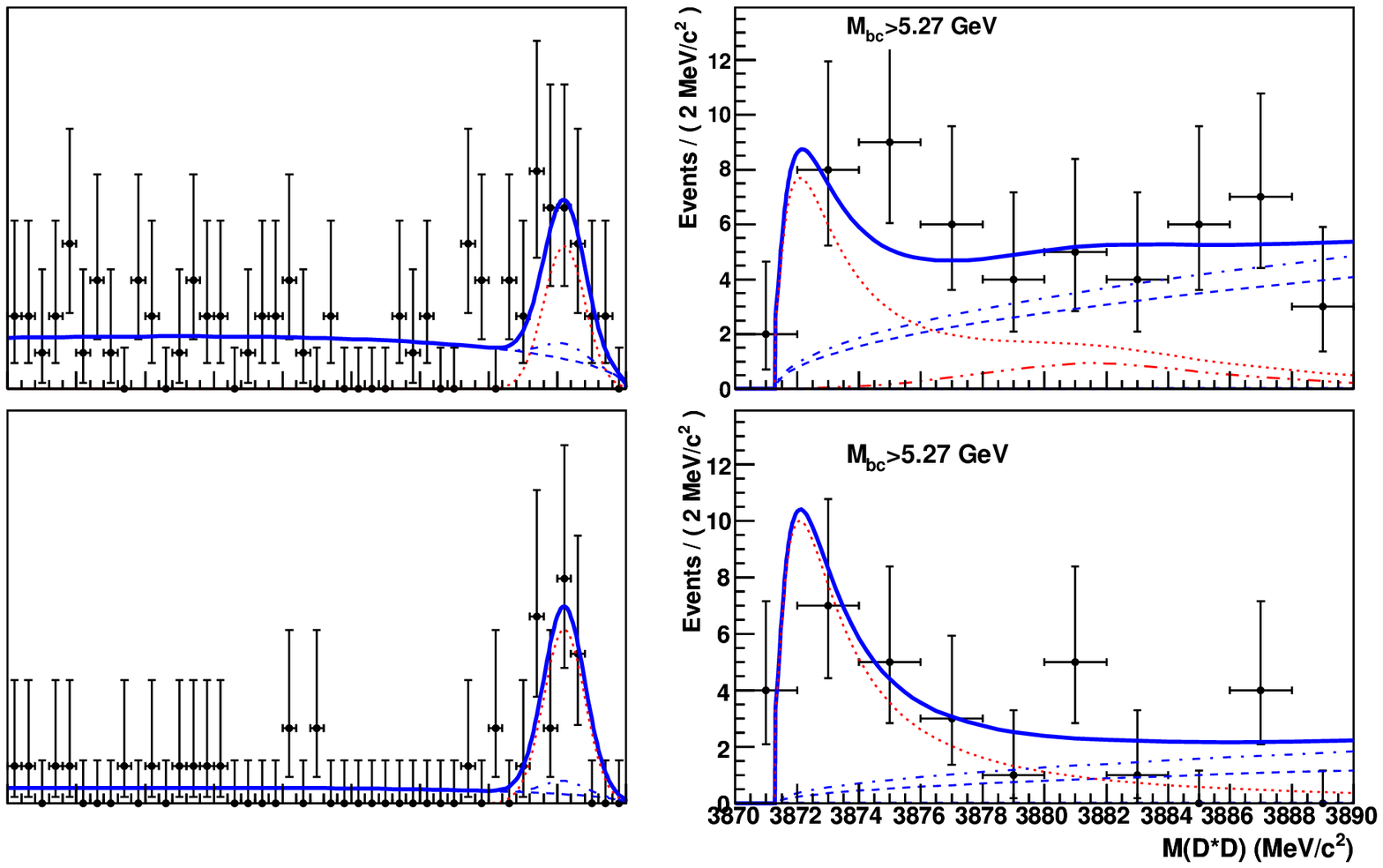}
\end{minipage}
\caption{Invariant $D^0\bar{D}^{*0}$ mass distribution measured by BaBar (left) and by Belle 
for $D^{*0} \rightarrow D^0\gamma$ (top right) and $D^{*0} \rightarrow D^0\pi^0$ (bottom right).}
\label{fig:X3872DDS}
\end{figure}

The decay $B^{+/0} \rightarrow X(3872)K^{+/0}$ with $X(3872) \rightarrow D^0\bar{D}^0\pi^0$
was first observed by Belle in 2006 with a significance of 6.4 standard deviations\ \cite{Gokhroo:2006bt}.
This observation was confirmed by BaBar last year\ \cite{Aubert:2007rva}.
In $B^{+/0} D^0\bar{D}^{*0}\rightarrow K^{+/0}$ decays, where the $\bar{D}^{*0}$ is reconstructed
via $\bar{D}^{0}\pi^0$ or $\bar{D}^{0}\gamma$, BaBar sees an excess near threshold
with a significance of 4.9$\sigma$ (see Fig.\ \ref{fig:X3872DDS} left).
A template fit is performed that takes into account threshold effects.
While the width of $\Gamma=3.0 ^{+1.9}_{-1.4}\mbox{(stat.)} \pm 0.9\mbox{(syst.)}$ MeV
is consistent with other measurements, the mass of 
$M=3875.1 ^{+0.7}_{-0.5}\mbox{(stat.)} \pm 0.5\mbox{(syst.)}$ MeV 
does not agree with of the mass of $M=3871.46 \pm 0.19$ MeV that is obtained
by averaging the measurements in the $X(3872) \rightarrow J/\psi\pi^+\pi^-$ mode.
This may raise the question whether the same particle is observed in both modes or
whether there are additional threshold effects.

Belle updated their analysis now including $D^{*0} \rightarrow D^0\gamma$ \ \cite{Adachi:2008su}.
Combining both $D^{*0}$ modes, a signal with a significance
of 8.8$\sigma$ is observed (see Fig.\ \ref{fig:X3872DDS} right).
A fit with a relativistic Breit-Wigner function and mass-dependent resolution yields
$M = 3872.6 ^{+0.5}_{-0.4}\mbox{(stat.)} \pm 0.4\mbox{(syst.)}$ MeV,
$\Gamma = 3.9 ^{+2.5}_{-1.3}\mbox{(stat.)} ^{+0.8}_{-0.3}\mbox{(syst.)}$ MeV, and
$\mathcal{B}(B \rightarrow X(3872)K)\times\mathcal{B}(X(3872) \rightarrow D^0\bar{D}^{*0}) 
= (0.73 \pm 0.17\mbox{(stat.)} \pm 0.08\mbox{(syst.)}) \times 10^{-4}$.
While the mass is closer to the average value in the $J/\psi\pi^+\pi^-$ mode,
some uncertainty due to the line shape model remains.
An alternative fit using a Flatt\'e function which assumes a mass below 
threshold\ \cite{Hanhart:2007yq} is able to describe the data, too.

\section{$\mathbf{X(3872) \rightarrow \psi\gamma}$}
Evidence for the decay $B^{+/0} \rightarrow X(3872)K^{+/0}$ with $X(3872) \rightarrow J/\psi\gamma$ 
was first seen in 2005 by Belle with a significance of 4.0$\sigma$\ \cite{Abe:2005ix} \
and one year later by BaBar with 3.4$\sigma$\ \cite{Aubert:2006aj}.
Now BaBar presented an updated analysis adding $B^{+/0} \rightarrow X(3872)K^{*+/0}$ events
and considering also $X(3872) \rightarrow \psi(2S)\gamma$ decays\ \cite{Aubert:2008rn}.

\begin{figure}
\centering
\begin{minipage}{0.45\textwidth}
\includegraphics[width=\textwidth]{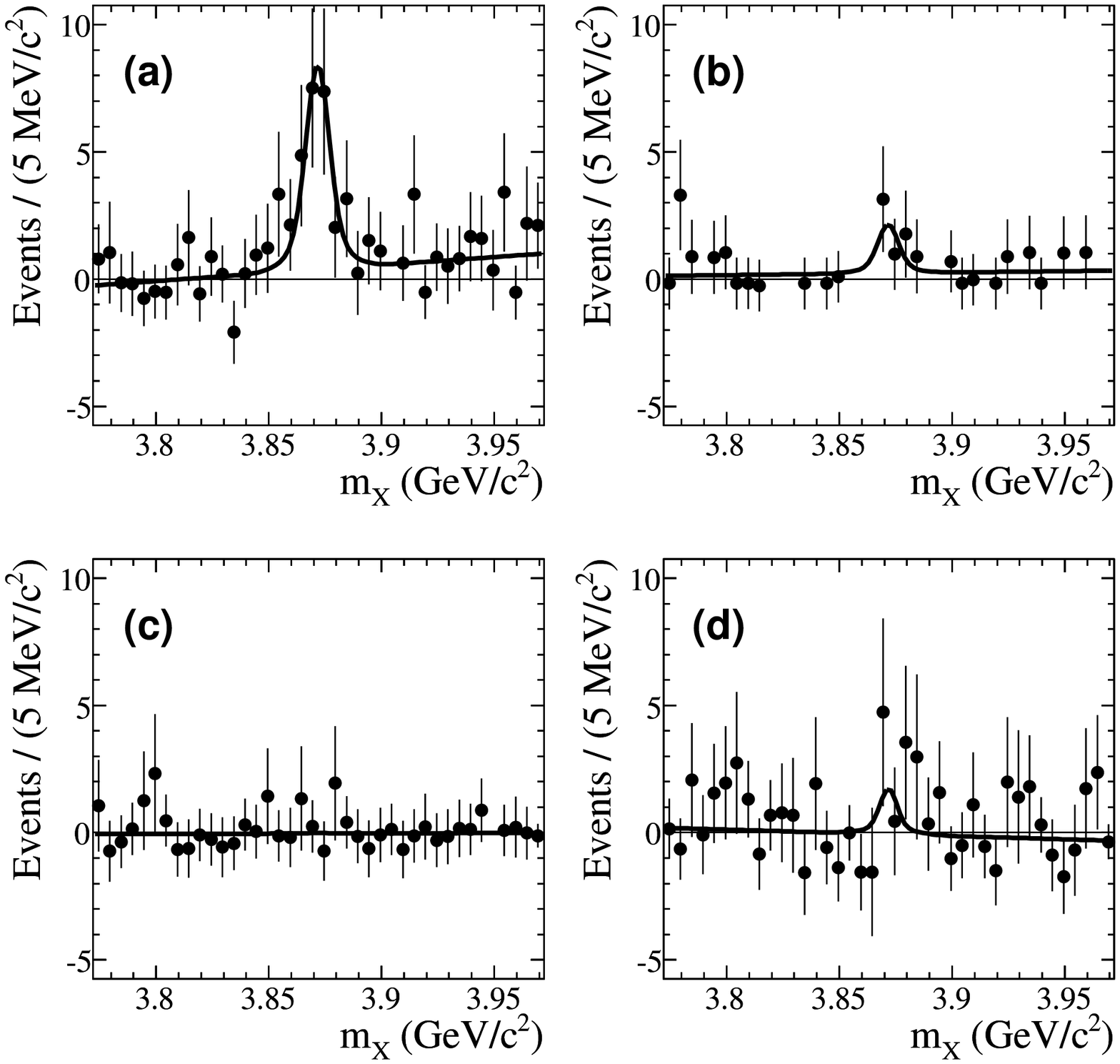}
\end{minipage}
\hspace*{0.05\textwidth}
\begin{minipage}{0.45\textwidth}
\includegraphics[width=\textwidth]{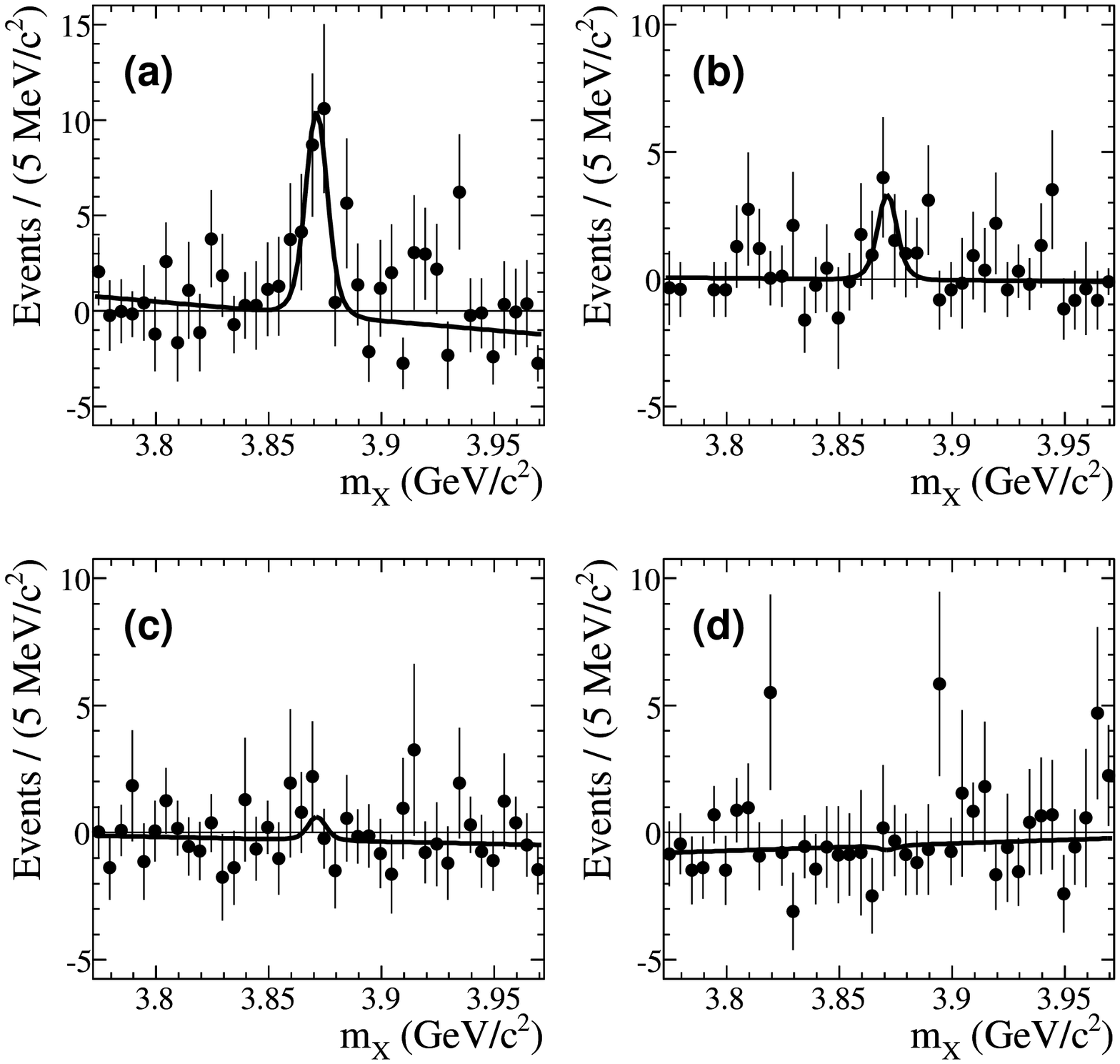}
\end{minipage}
\caption{Invariant $J/\psi\gamma$ (left) and $\psi(2S)\gamma$ (right) mass distribution measured by BaBar.
The four plots for each mode correspond to $B$ decays channels with $K^+$ (a), $K^0_s$ (b), $K^{*+}$ (c), 
and $K^{*0}$ (d).}
\label{fig:X3872psigamma}
\end{figure}

Signals with a significance of 3.6$\sigma$ and 3.5$\sigma$ are seen in the $J/\psi\gamma$
and $\psi(2S)\gamma$ modes, respectively (see Fig.\ \ref{fig:X3872psigamma}).
For the $B$ decays to $\psi K^+$ branching ratios of
$\mathcal{B}(B^+ \rightarrow X(3872)K^+)\times\mathcal{B}(X(3872) \rightarrow J/\psi\gamma) 
= (2.8 \pm 0.8\mbox{(stat.)} \pm 0.1\mbox{(syst.)}) \times 10^{-6}$ and
$\mathcal{B}(B^+ \rightarrow X(3872)K^+)\times\mathcal{B}(X(3872) \rightarrow \psi(2S)\gamma) 
= (9.5 \pm 2.7\mbox{(stat.)} \pm 0.6\mbox{(syst.)}) \times 10^{-6}$ are measured.
Limits for the other three $B$ decay modes where no significant signal is observed are quoted.
The determined relative branching ratio of 
$\mathcal{B}(X(3872) \rightarrow \psi(2S)\gamma)/\mathcal{B}(X(3872) \rightarrow J/\psi\gamma)
= 3.4 \pm 1.4$ is orders of magnitude higher than expected in $D^0\bar{D}^{*0}$ 
molecule models\ \cite{Swanson:2004pp}.

\section{$\mathbf{Z(4430)^+ \rightarrow \psi\pi^+}$}
In 2008 Belle published the observation of a resonance in the $\psi(2S)\pi^+$ invariant mass
spectrum in $B^{+/0} \rightarrow \psi(2S)\pi^+ K^{0/-}$ decays with a significance of
6.5$\sigma$\ \cite{Choi:2007wga}.
In this analysis events with a $K\pi$ mass consistent with a $K^*(890)$ or $K^*(1430)$ were
removed and the resulting $M(\psi(2S)\pi^+)$ distribution was fitted.
From the fit the mass $M = 4433 \pm 4\mbox{(stat.)} \pm 2\mbox{(syst.)}$ MeV, the width
$\Gamma = 45 ^{+18}_{-13}\mbox{(stat.)} ^{+30}_{-13}\mbox{(syst.)}$ MeV, and the branching
ratio $\mathcal{B}(B \rightarrow Z^+ K) \times \mathcal{B}(Z^+ \rightarrow \psi(2S)\pi^+)
= (4.1 \pm 1.0\mbox{(stat.)} \pm 1.4\mbox{(syst.)}) \times 10^{-5}$ were extracted.

BaBar performed a detailed study of $B^{+/0} \rightarrow J/\psi\pi^+ K^{0/-}$ and
$B^{+/0} \rightarrow \psi(2S)\pi^+ K^{0/-}$ decays\ \cite{Aubert:2008nk}.
As the most prominent features in the Dalitz plot are the $K^*(890)$ and $K^*(1430)$
resonances a good understanding of the $K\pi$ mass spectrum is important.
BaBar is able to describe it well with $S$-, $P$-, and $D$-waves.
This composition and the angular distributions of the components have a strong
influence on the $\psi\pi^+$ mass spectrum.
On the other hand they cannot create narrow peaks in the $M(\psi\pi^+)$ distribution.

\begin{figure}
\centering
\begin{minipage}{0.67\textwidth}
\includegraphics[width=\textwidth]{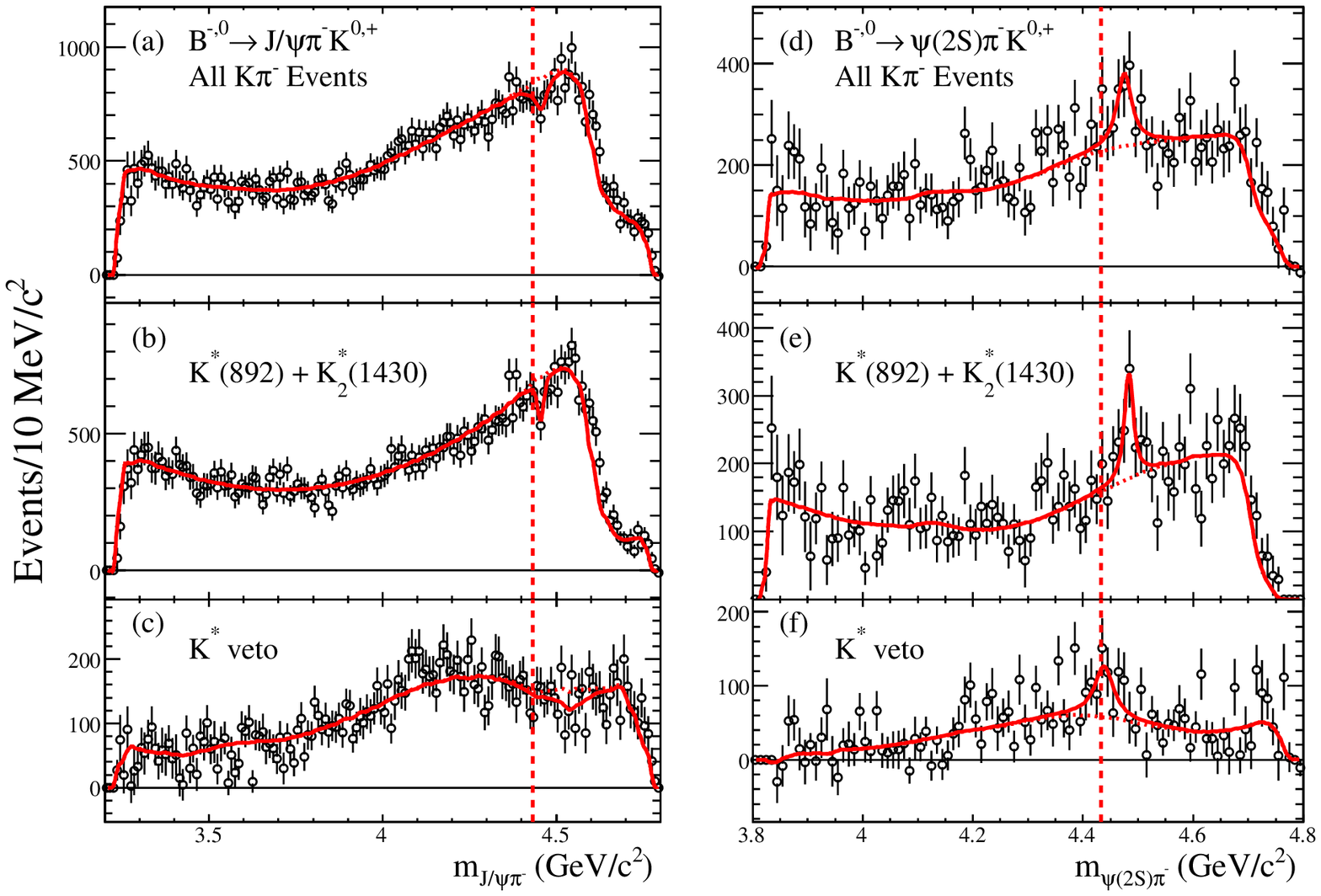}
\end{minipage}
\begin{minipage}{0.32\textwidth}
\includegraphics[width=\textwidth]{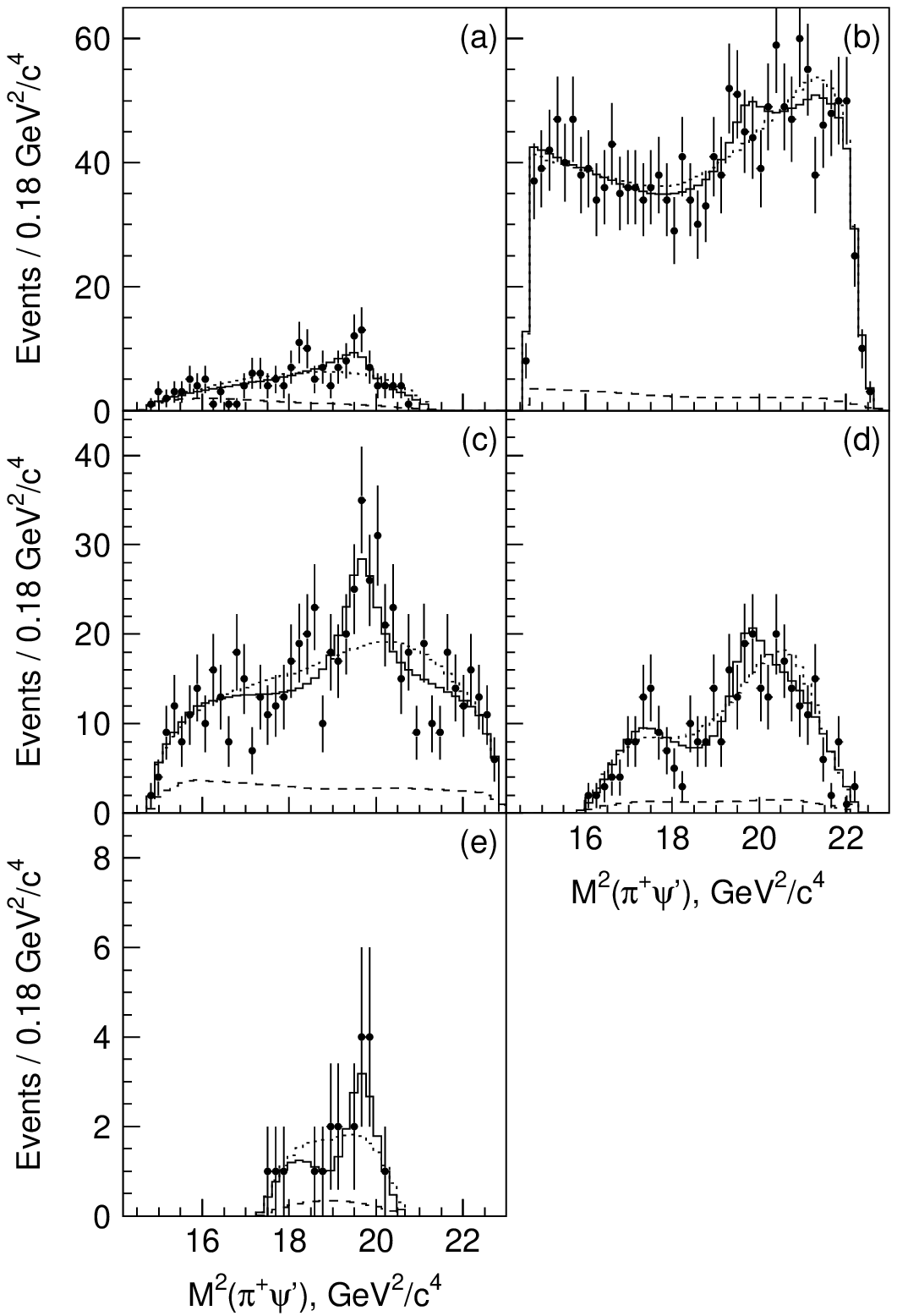}
\end{minipage}
\caption{Invariant $J/\psi\pi^+$ and $\psi(2S)\pi^+$ mass distributions for
all events and events inside and outside the $K^*(890)$ and $K^*(1430)$ regions
measured by BaBar (left).
Invariant $\psi(2S)\pi^+$ mass distributions in different regions of $M(K\pi)$
measured by Belle (right).}
\label{fig:Zpsipi}
\end{figure}

BaBar looks at the $J/\psi\pi^+$ and $\psi(2S)\pi^+$ mass distributions for
all events, events in the $K^*(890)$ and $K^*(1430)$ region, and events outside the $K^*$ 
regions and compares them to the projections of the $K\pi$ components (see Fig.\ \ref{fig:Zpsipi} left).
No signal is seen in the $M(J/\psi\pi^+)$ spectra and no significant signal
in the $M(\psi(2S)\pi^+)$ spectra.
For the total sample there is an excess of 2.7$\sigma$ in the $\psi(2S)\pi^+$ mass distribution 
at $M=4476 \pm 8\mbox{(stat.)}$ MeV with $\Gamma=32 \pm 16\mbox{(stat.)}$ MeV.
Using the values measured by Belle a 95\% confidence level limit on the branching ratio
of $\mathcal{B}(B \rightarrow Z^+ K) \times \mathcal{B}(Z^+ \rightarrow \psi(2S)\pi^+)
< 3.1 \times 10^{-5}$ is set.
A comparison of the background subtracted $M(\psi(2S)\pi^+)$ distribution with the one
published by Belle shows good statistical agreement.

Belle reanalyzed their data with a fit to the Dalitz plot taking into account all known 
$K\pi$ resonances\ \cite{Belle:ZDalitz}.
The fit with $Z^+$ resonance included (solid line in Fig.\ \ref{fig:Zpsipi} right) 
describes the data significantly better than without (dotted line)
by 6.4$\sigma$.
Systematic variation of the fit model yield a significance of 5.4$\sigma$.
The updated measurements of the $Z^+(4430)$ properties are 
$M=4443^{+15}_{-12}\mbox{(stat.)} ^{+19}_{-13}\mbox{(syst.)}$ MeV,
$\Gamma=107 ^{+86}_{-43}\mbox{(stat.)} ^{+74}_{-56}\mbox{(syst.)}$ MeV, and
$\mathcal{B}(B \rightarrow Z^+ K) \times \mathcal{B}(Z^+ \rightarrow \psi(2S)\pi^+)
= (3.2 ^{+1.8}_{-0.9}\mbox{(stat.)} ^{+5.3}_{-1.6}\mbox{(syst.)}) \times 10^{-5}$.

\section{$\mathbf{Z(4050/4250)^+ \rightarrow \chi_{c1}\pi^+}$}
\begin{figure}
\centering
\begin{minipage}{0.49\textwidth}
\includegraphics[width=\textwidth]{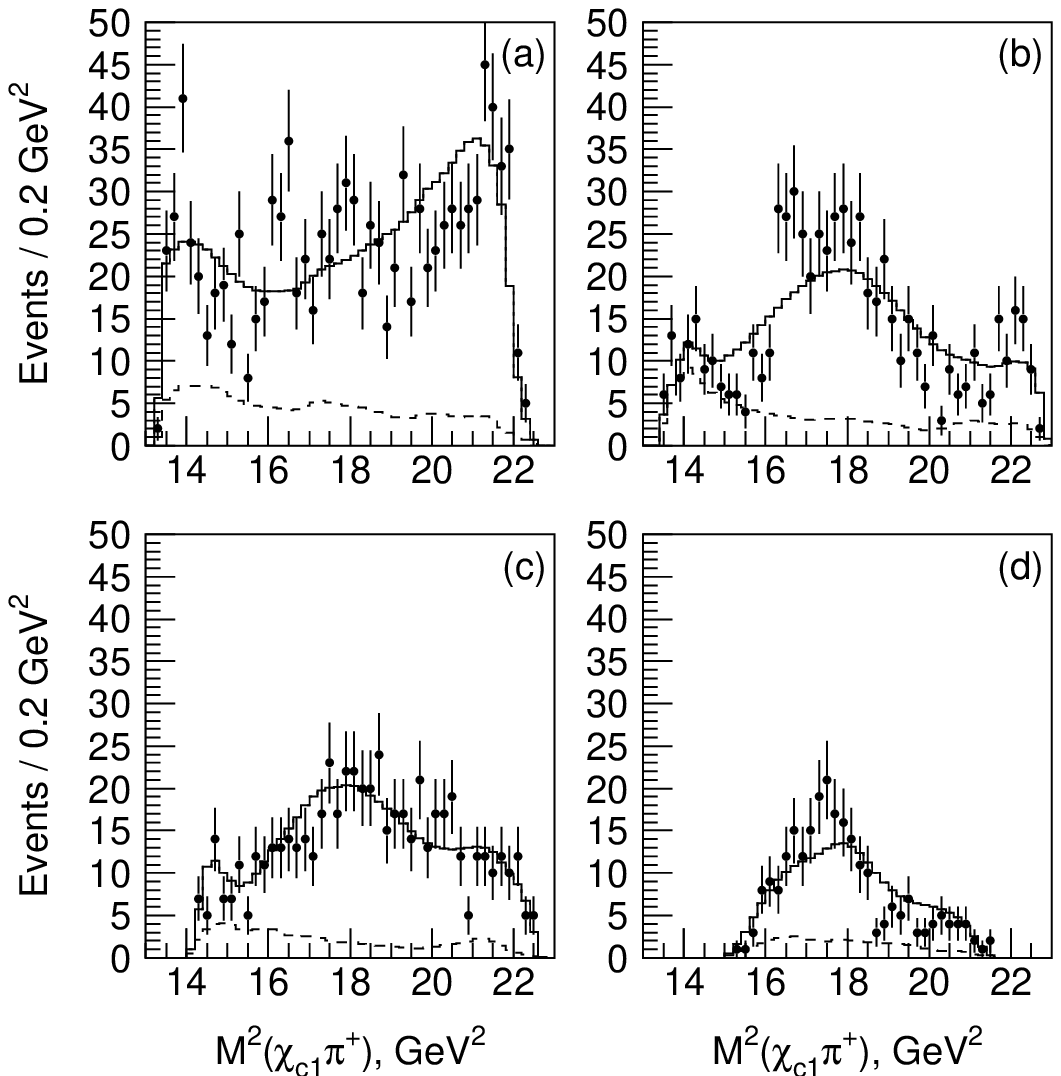}
\end{minipage}
\begin{minipage}{0.49\textwidth}
\includegraphics[width=\textwidth]{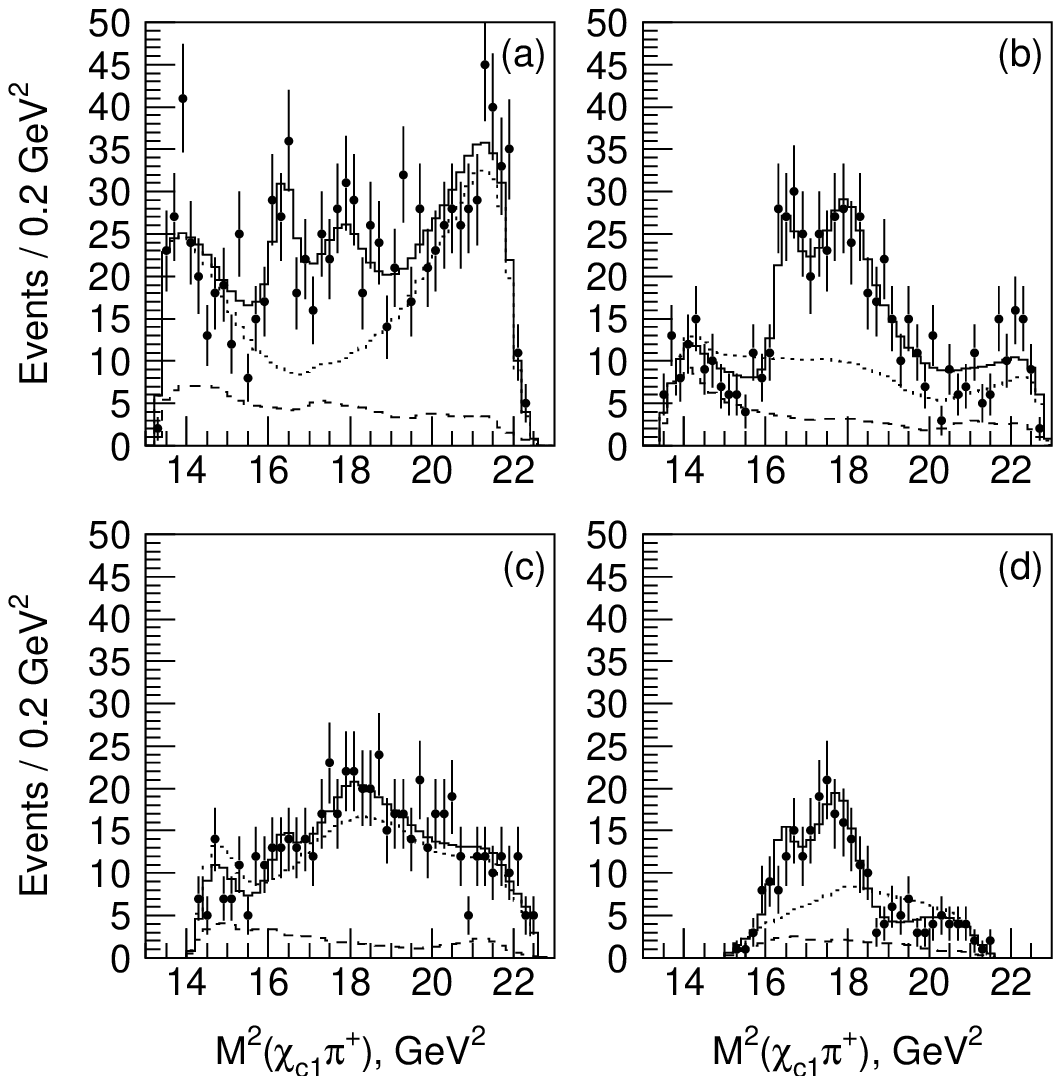}
\end{minipage}
\caption{Invariant $\chi_{c1}\pi^+$ mass distributions for different $M(K\pi)$ regions
measured by Belle with fit projection assuming no $Z^+$ state (left) and two $Z^+$ states (right).}
\label{fig:Zchic1pi}
\end{figure}

As the first candidate for a charged exotic state was observed as a $\psi(2S)\pi^+$
resonance, this suggests to search for similar charged states in combinations of other
charmionia with pions.
In a Dalitz plot analysis of $\bar{B}^0 \rightarrow \chi_{c1}\pi^+K^-$ events including
all known $K\pi$ resonances Belle observes two new states, called $Z(4050)^+$
and $Z(4250)^+$\ \cite{Mizuk:2008me} (see Fig.\ \ref{fig:Zchic1pi}).
From the change in the fit likelihood value a significance of the two-$Z^+$-state
versus the no-$Z^+$-state hypothesis of 13.2$\sigma$ is determined.
Taking into account systematic variations of the fit model the significance is
at least 8.1$\sigma$.
Compared to the one-$Z^+$-state hypothesis the two states are favored by 5.7$\sigma$,
or 5.0$\sigma$ including systematics.
The masses, widths and branching ratios of the two new states are measured to be 
$M(Z(4050)^+) = 4051 \pm 14\mbox{(stat.)} ^{+20}_{-41}\mbox{(syst.)}$ MeV,
$M(Z(4250)^+) = 4248 ^{+44}_{-29}\mbox{(stat.)} ^{+180}_{-35}\mbox{(syst.)}$ MeV,
$\Gamma(Z(4050)^+) = 82 ^{+21}_{-17}\mbox{(stat.)} ^{+47}_{-22}\mbox{(syst.)}$ MeV,
$\Gamma(Z(4250)^+) = 177 ^{+54}_{-39}\mbox{(stat.)} ^{+316}_{-61}\mbox{(syst.)}$ MeV,
$\mathcal{B}(\bar{B}^0 \rightarrow Z(4050)^+ K^-) \times \mathcal{B}(Z(4050)^+ \rightarrow \chi_{c1}\pi^+)
= \linebreak (3.0 ^{+1.8}_{-0.8}\mbox{(stat.)} ^{+3.7}_{-1.6}\mbox{(syst.)}) \times 10^{-5}$, and
$\mathcal{B}(\bar{B}^0 \rightarrow Z(4250)^+ K^-) \times \mathcal{B}(Z(4250)^+ \rightarrow \chi_{c1}\pi^+)
= (4.0 ^{+2.3}_{-0.9}\mbox{(stat.)} ^{+19.7}_{-0.5}\mbox{(syst.)}) \times 10^{-5}$.

\section{Conclusions and outlook}
The study of the decay $X(3872) \rightarrow D^0\bar{D}^{*0}$ is an interesting topic
because it is very close to threshold.
While the $X(3872)$ mass measured by BaBar is significantly higher than the average
mass measured in the $J/\psi\pi^+\pi^-$ mode, the latest Belle measurement agrees with it
within 2$\sigma$.
A detailed study of the exact line shape may reveal important information about the
nature of the $X(3872)$, but requires a much larger data sample.
The first observation of the decay $X(3872) \rightarrow \psi(2S)\gamma$ and the
measured high branching ratio with respect to the $J/\psi\gamma$ mode disfavor
the hypothesis of a $D^0\bar{D}^{*0}$ molecule.

The first observation of a charged charmonium-like state, the $Z(4430)^+$,
was confirmed in a Dalitz plot analysis by Belle.
But so far this particle is not confirmed by any other experiment.
BaBar does not see a significant signal
in the $J/\psi\pi^+$ and $\psi(2S)\pi^+$ mass spectra.
Although statistical agreement with the Belle data is established,
the conclusions of both experiments differ.
Two further charged states with even higher significance have been observed
as $\chi_{c1}\pi^+$ resonances by Belle.
A clear picture of the $Z^+$ states with precise measurements of their properties
may only arise at upgraded B-factories.


\begin{footnotesize}



%

\end{footnotesize}


\end{document}